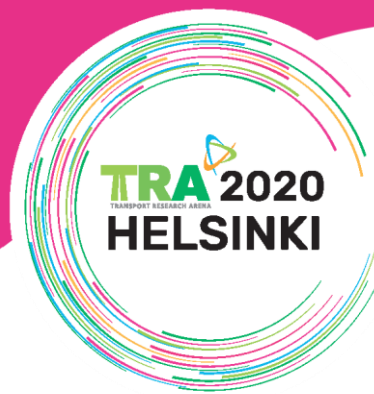

# An analysis of European crash data and scenario specification for heavy truck safety system development within the AEROFLEX project

Ron Schindler[a]*, Michael Jänsch[b], Heiko Johannsen[b], András Bálint[a]

[a]*Department of Mechanics and Maritime Sciences Chalmers University of Technology, Göteborg, Sweden*
[b]*Accident Research Unit, Medizinische Hochschule Hannover, Hannover, Germany*

**Abstract**

Heavy goods vehicles (HGVs) are involved in 4.5% of police-reported road crashes in Europe and 14.2% of fatal road crashes. Active and passive safety systems can help to prevent crashes or mitigate the consequences but need detailed scenarios to be designed effectively. The aim of this paper is to give a comprehensive and up-to-date analysis of HGV crashes in Europe. The analysis is based on general statistics from CARE, results about trucks weighing 16 tons or more from national crash databases and a detailed study of in-depth crash data from GIDAS. Three scenarios are identified that should be addressed by future safety systems: (1) rear-end crashes with other vehicles in which the truck is the striking partner, (2) conflicts during right turn maneuvers of the truck and a cyclist and (3) pedestrians crossing the road perpendicular to the direction of travel of the truck.

*Keywords:* heavy goods vehicle; crash scenarios; GIDAS; CARE; crash causation; European national crash data

* Corresponding author. Tel.: +46 70 509 1536;
*E-mail address:* ron.schindler@chalmers.se
https://orcid.org/0000-0002-5895-3387

# 1. Background

In the year 2015, more than 1 million crashes happened on European roads, out of which around 24,000 resulted in fatalities. Heavy goods vehicles (HGV) were involved in 4.5% of all crashes and 14.2% of fatal crashes, indicating an overrepresentation of HGV involvement in fatal crashes (Source: CARE, 2019). Development and implementation of active and passive safety systems in HGVs can help to avoid or mitigate these crashes, and the basis for the design of these systems is a deep understanding of the underlying mechanisms and factors influencing the crash causation and outcome.

Concerning the general crash situation for HGVs, Zhu and Srinivasan (2011) analyzed data from the Large Truck Crash Causation Study (LTCCS, USA) and concluded that head-on collisions and collisions at intersections are the most serious crash types in terms of the overall injury outcome. Furthermore, in crashes that involve heavy trucks, fatally injured persons are usually occupants of the crash opponent vehicles. Khorashadi et al. (2005) looked at driver injury severity in crashes between cars and heavy trucks, identifying factors that are associated with an increase or decrease of the risk for serious injuries such as crash type and time of the day. The authors conclude that the influence of these factors is especially noteworthy as the magnitude varies substantially depending on crash location (i.e. if the crash happens in an urban or rural area). US data from Woodroofe and Blower (2015) identified rollover and head-on collisions as the main collision types for truck driver injuries, accounting for 73% of serious and fatal injuries of truck drivers. Kockum et al. (2017) analyzed European crash data and concluded that HGV occupants are injured in 10-20% of HGV crashes, while the corresponding figures are 50-55% for car occupants and 30-35% for vulnerable road users (VRU), supporting previous findings that car occupants and VRUs comprise the largest group of casualties in truck related crashes.

Vulnerable road users, lacking a protective shell (e.g. crumple zone, airbag), may be especially exposed in crashes with HGV involvement. Kim et al. (2007) investigated factors influencing the injury severity of cyclists in the US and correlate the involvement of a truck in the crash with a significant increase in the likelihood of a fatal injury. Lee and Abdel-Aty (2005) analyzed injury risk for pedestrians at intersections with respect to different vehicle types and conditions in Florida (USA). The authors conclude that passenger cars were involved in more crashes than trucks – however, the larger size of HGVs was correlated to an increased likelihood of severe injuries. Adminaite et al. (2015) describe crashes between trucks and VRUs especially problematic due to the vehicle size and difference in mass and indicate that the main reason for these crashes is the problematic field of view for truck drivers, making VRUs particularly prone to be in the blind spot and overseen by the truck driver.

Rezapour et al. (2018) analyzed transport data from Wyoming (USA) and conducted a violation analysis. The authors conclude that in more than 80% of all crashes involving a truck, the truck drivers are the party at fault, which emphasizes the need to introduce more safety systems into trucks. Seiniger et al. (2015) have identified that the development of new safety systems of trucks for cyclist protection should be focused on right turning maneuvers and propose a test methodology to validate new active safety systems. Further details of crash scenarios classified by injured road users are provided in the already referenced technical report by Kockum et al. (2017).

In 2008, Knight et al. identified a lack of robust European crash data especially for large trucks. Overall, the literature review of results for crashes involving heavy trucks has revealed various limitations, especially regarding the study of crash causation. Sandin et al. (2014) identified a lower crash rate for extra-long trucks that are allowed on Swedish roads (longer than 18.75m, ~62ft), but this analysis is limited to Swedish data and no information of crash causation is available. Evgenikos et al. (2016) have analyzed crashes in Europe involving heavy trucks based on European data from the Community Database on Accidents on the Roads in Europe (CARE) and in-depth data from the SafetyNet project. The authors conduct an accident causation analysis, but focus is put on injury outcome and no pre-crash information (e.g. initial speed, environment conditions) is available.

Furthermore, the limited amount of information from the literature usually includes different types of trucks, i.e. all vehicles with a weight above 3.5t (~8,000lb). However, there are significant differences in vehicle types in this category, that could range from heavy vans like a Mercedes Sprinter to long-haul truck-trailer combinations such as the Volvo FH. Different types of HGVs have different characteristics (e.g. vehicle dynamics, field of view), and a more detailed classification would be important for a more directed safety system development. In addition, previous studies typically focus on injury outcome of crashes, but miss describing the circumstances of the crashes (e.g. weather condition, road condition).

Moreover, several studies are based on data from North America. Wang and Wei (2016) showed in their analysis that benefits achieved by active safety systems in one country cannot easily be transferred to other countries (or even continents), emphasizing the need to analyze regional data. Due to different road infrastructure designs (e.g. wider lanes) and vehicle designs (e.g. conventional cab design in North America compared to flat nose design in Europe), the representativeness of study results obtained in North America to the situation in Europe are limited.

The aim of this paper is to give a comprehensive and up-to-date analysis of HGV crashes in Europe, focusing on heavy long-haul trucks with a combination weight above 16t (further referred to as 16t+ trucks). The results in this paper summarize and extend the corresponding analysis in the AEROFLEX project (AEROdynamic and FLEXible Trucks for Next Generation of Long Distance Road Transport) reported in deliverable D5.1 (Schindler et al. 2018).

## 2. Method

The approach to identify relevant crash scenarios consists of three levels. First, data from the Community Database on Accidents on the Roads in Europe (CARE) is extracted to get a general understanding of crashes with the involvement of heavy goods vehicles (3.5t+ trucks) in Europe. As CARE contains high level data, this first level of the analysis needs to be complemented by other data sources for the identification of relevant crash scenarios. More detailed information can be obtained from national crash databases, which is the second level of the analysis. These contain more information, such as vehicle weight, that allow for a filtering of crashes to include only those with the involvement of a 16t+ truck. Specifically, national data from Sweden, Spain and Italy is analyzed.

The third level of analysis is to look at in-depth data from the German In-Depth Accident Study (GIDAS). In-depth crash databases contain detailed crash scenario description, crash reconstruction and information about causation factors. At the same time, their sampling region and case count are substantially smaller compared to CARE or the national databases. Consequently, each of the three levels of the analysis provides information that complements results of the other two and a full picture can only be obtained by their combination.

The analysis addresses police-reported crash severity levels defined as follows:

- Fatal crash: A crash in which at least one person was fatally injured (the person died from the crash within 30 days). Other persons may be involved, and their potential injuries might range from slight injuries to fatal injuries.
- Severe crash: A crash in which at least one person was severely injured (hospitalized for at least 24 hours), and nobody was fatally injured. Other persons may be involved, and potential injuries might range from slight injuries to severe injuries.
- Slight crash: A crash in which at least one person was slightly injured (hospitalized for less than 24 hours or not hospitalized), and nobody was fatally or severely injured. Other persons without injuries or with slight injuries may be involved.

An additional crash severity level for the analysis is "KSI crashes" which are defined as crashes in which at least one person was killed or severely injured; equivalently, KSI crashes are the union of fatal and severe crashes.

### 2.1. CARE

As indicated above, EU crash data on an aggregated level is available in CARE. This database is used to obtain general estimates from police-reported crash data from all EU member states (EC, 2018a). The set of variables in CARE is specified in the CaDaS glossary (EC, 2018b). As vehicle weight is not included as a variable, all results based on CARE queries will be using the general definition of HGVs, i.e. with a total weight of 3.5t or greater.

Data from 2010 to 2015 from EU28 was used for this analysis. In order to identify crashes involving trucks, the dataset is filtered for crashes involving the traffic element groups "Goods vehicle", "Goods vehicle with a total weight above 3,5t" and "Road tractors". If there were gaps in the data (e.g. years without reported data for a given country), those have been filled by the data value from the same country for the closest year, using data from the earlier year in case of two years that were equally close. Furthermore, for years where all crashes were coded in the category "unknown", the crashes were distributed to the available categories by interpolating the total number of crashes with the distribution from the closest year.

Police-reported injury severity levels as defined in the previous section are considered for the analysis. Italy, Finland and Estonia do not distinguish between severe and slight injuries, but rather report a generic number of injured persons. Therefore, for these countries, it was assumed that 14% of all reported non-fatal injuries were serious and 86% were slight in order to get an approximation. The estimation of 14% used here is based on a study of Italian data, see the next section, because Italy gives the large majority of cases from the above three countries.

### 2.2. National crash databases

In the second stage of the analysis, national crash databases have been analyzed. The Swedish national crash database (STRADA, see Transportstyrelsen (2019) for more information) was analyzed for 16t+ trucks from the years 2000 to 2016. Spanish national crash data (Dirección General de Tráfico, 2019) was analyzed for the years 2014 to 2016 for 16t+ trucks. Italian national crash data (Istat, 2019) was analyzed for the years 2010 to 2016. In 2015, these three countries together accounted for 21% of traffic fatalities in Europe, with Italy and Spain also being amongst the highest populated countries in Europe.

Since Italian data does not provide a distinction between severe and slight injuries, the reported numbers of non-fatal injured persons were distributed to slight and severe injuries according to a study by the Italian road infrastructure administration (Ministero delle Infrastrutture e dei Trasporti, 2010), with 86% of the injuries being recoded as slight and the remaining 14% as severe. This distribution also corresponds to other European countries which have a range of 12%-16% of seriously injured persons as a percentage of all injuries, according to CARE.

Additionally, Italian data did not include information about light conditions, hence the corresponding analysis will be given for the other two countries only.

Beyond the analysis of existing data, this study included a harmonization and analysis of the basic crash types. This analysis required a re-coding of data in the Italian and Spanish databases to make them comparable with the Swedish classification.

*2.3. GIDAS*

For the third stage, in-depth crash data involving 16t+ trucks from GIDAS was analyzed. GIDAS is the largest in-depth road accident study in Germany, where about 2,000 crashes from the areas of Hannover and Dresden are investigated annually. The crash investigation teams are alarmed by the police and go on-scene to crashes with at least one injured person. Up to 3,000 variables are recorded per crash, including vehicle data, crash information, road design, active and passive safety systems, crash scene details and crash causes. After the data collection, each accident is reconstructed to obtain information on crash kinematics and sequences. Due to a statistical sampling process, the collected crashes are suitable of representing the German crash situation (Johannsen et al., 2017).

The accident type catalog developed by the German insurance association GDV (see Institut für Straßenverkehr, 1998) is used for this study which provides more detailed information than the crash types in the national data analyzed before. It describes the initial conflict situation of the crash (which is not necessarily the way the involved vehicles collided). There are over 200 specific crash types which are categorized in 7 main categories.

The opponent that caused the most severe injury to the truck driver was chosen as the collision partner of the trucks for further analysis. If there was no most severe injury available (e.g. the person was uninjured), then the partner of the collision which caused the most damage to the truck was chosen. If damage was not available or visible, then the partner of the first collision was chosen.

## 3. Results

This section describes the results from all three analysis levels. For the percentages reported in this section, injuries and categorical factors that were coded as "other" or "unknown" were excluded if not indicated otherwise.

*3.1. CARE*

The high-level analysis based on CARE data gives a general characterization of crashes involving 3.5t+ trucks in Europe, see Fig. 1. The figure shows the names of environmental variables, with the most frequent values specified between parentheses, with the corresponding prevalence of these values indicated by the length of the bars. For example, for "Weather" the most common value is "Dry/clear" which is present in 81% of crashes, followed by "Rain", present in 11% of crashes, and other values of the variable, present in 7% of the crashes. The results indicate that most crashes involving 3.5t+ trucks in EU-28 occur in dry/clear weather (81%), daylight (78%), on roads that are not highways (77%), on roads with a dry surface (72%) and in rural areas (57%).

The analogous results for crashes with a severe or fatal outcome show that KSI crashes involving HGVs in EU-28 can be characterized by dry/clear weather (82%), not-highway roads (77%), daylight (73%), dry road surface (72%) and rural environment (65%). In other words, the results are generally similar to those for all injury crashes, but with greater percentages of rural crashes (65% vs 57%) and crashes in darkness (22% vs 18%).

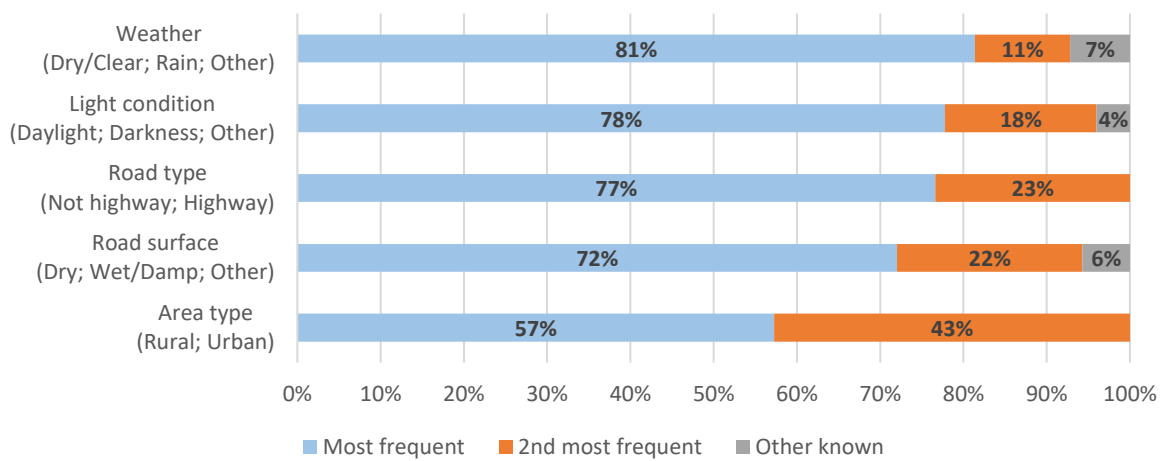

Fig. 1 Characterization of crashes involving HGVs in Europe in terms of environmental variables, based on CARE

Additional analysis was performed identifying that people injured in HGV crashes in EU-28 are mainly car occupants (55%), followed by HGV occupants (21%), vulnerable road users (16%) and other road users (8%). The distribution of injured road users by road user type and age is provided in Table 1 below, showing that the largest group of injured people in HGV crashes are car occupants and HGV occupants between 25 and 64 years. This age group is especially prevalent for HGV occupants, as 85% of all injured HGV occupants and 86% of all KSI HGV occupants are in this group. Notably, the percentage of young (<25 years) and old (>64 years) VRUs is much higher among KSI road users than for all injury levels. The gender distribution, provided in Fig. 2 below, indicates that males are more frequently injured in crashes with HGV involvement than females, with males having a total share of 65% of all injuries and 71% of KSI injuries. While the gender distribution is more balanced for injured car occupants, it is very skewed towards males for injured HGV occupants (92% of whom are males).

All results based on CARE data use the most inclusive definition of heavy goods vehicles, based on a total weight of 3.5t or higher. As indicated in Section 2, these results will be complemented with analysis of national crash data and in-depth crash data that allow for filtering on combination weight. For the rest of this section, all results will be addressing 16t+ trucks, unless indicated otherwise.

Table 1. Joint distribution of age and road user group for people injured in crashes with HGV involvement in EU-28, separately for all injury levels and fatally or severely injured (KSI) people, rounded values, based on CARE

|  |  | <18 | 18-24 | 25-64 | >64 | TOTAL |
|---|---|---|---|---|---|---|
| All injured road users | HGV occupant | 0% | 2% | 18% | 1% | 21% |
|  | Car occupant | 4% | 9% | 36% | 6% | 55% |
|  | VRU | 2% | 2% | 9% | 3% | 16% |
|  | Other | 1% | 1% | 5% | 1% | 8% |
|  | TOTAL (n=462,107) | 7% | 14% | 69% | 10% | 100% |
| KSI road users | HGV occupant | 0% | 2% | 17% | 1% | 20% |
|  | Car occupant | 3% | 8% | 30% | 8% | 48% |
|  | VRU | 3% | 3% | 14% | 6% | 25% |
|  | Other | 1% | 1% | 5% | 1% | 8% |
|  | TOTAL (n=109,825) | 6% | 13% | 66% | 16% | 100% |

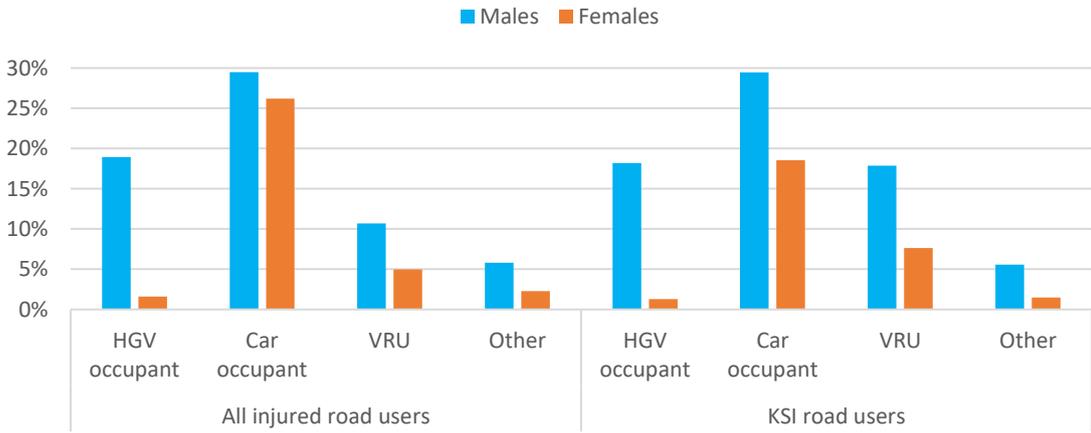

Fig. 2 Distribution of injured, respectively KSI, road users in crashes in EU-28 with HGV involvement, based on CARE

### 3.2. National crash databases

The national crash statistics for 16t+ trucks from the analyzed countries (Sweden, Spain, Italy) generally follow the trends observed in CARE. Most crashes occur in dry/clear weather conditions (SWE 77%, ESP 88%, ITA 76%) during daylight conditions (SWE 73%, ESP 74%, ITA n.a.). Bigger differences between countries (and towards CARE) can be observed for surface conditions, where most crashes occur with dry surfaces (SWE 51%, ESP 83%, ITA 81%) and on roads that are not highways (SWE 81%, ESP 54%, ITA 69%) outside city areas (SWE 60%, ESP 87%, ITA 66%). Furthermore, crash type categorization is available from national crash databases and can be compared across countries after re-coding as described in section 2.2.

Across all analyzed countries and across all crash scenarios, fatal crashes have the smallest share, whereas slight injury crashes are the majority. In Sweden, 86% of rear-end crashes are slight crashes, whereas the crash type including the highest percentage of fatal crashes are VRU accidents of which 16% have a fatal injury outcome. Also in Spain, rear-end crashes have the highest share of slight injury crashes among the known accident types. However, meeting/overtaking crashes show the highest share of fatal crashes (18.3%), which is similar to the share of severe crashes (18.7%) in this crash type. VRU crashes are also critical, having the second highest share of fatal (12%) and severe (16%) crashes within a crash type. In Italy, the injury share of VRU related crashes is comparably high as in Spain. Furthermore, the share of injuries of single vehicle accidents is very low (3%), whereas the injury outcome of meeting/overtaking (32%), rear-end (31%) and intersection crashes (26%) is very similar to each other. In comparison to Sweden and Spain, the fatality share in meeting/overtaking crashes is similar, whereas the number of slight injuries is two to four times higher in Italy. A similar tendency (i.e. a higher share of slight crashes compared to the other two countries) can be seen in the Italian data for intersection accidents.

*3.3. GIDAS*

For the GIDAS analysis, cases from the years 2000-2017 were used, for which 1,091 trucks with a gross vehicle weight above 16t were available in the data at the time of analysis. The general analysis of all 16t+ trucks in injury accidents revealed that the majority of crashes happened during daytime (75%) and most crashes had occurred outside city limits (59%) and on motorways (42%). Most trucks had an accident outside of junctions, either on a straight stretch of road (55%) or in a curve (9%), which corresponds to the high share of motorway accidents. Crashes at junctions (11%) or crossings (20%) were less frequent. The analysis of the main accident type categories within the analyzed GIDAS sample shows that nearly half of the crashes (49.0%) occurred due to a conflict with another vehicle in longitudinal traffic (e.g. head-on or rear-end collisions). Accidents when turning off the priority road (13.2%) or when entering or crossing the priority road (13.3%) account for slightly over 26% of cases.

For an analysis of the type of road user as crash opponent, cases with unknown accident type or unknown collision partner as well as single vehicle crashes were discarded. A summary of the types of collision partners of 16t+ trucks and crash type frequencies can be found in Table 2, showing that 16t+ trucks are mostly involved in crashes with cars (44.6%), the majority of them occurring due to a conflict in longitudinal traffic (in 55.9% of truck-to-car cases, see Fig. 3). Of these crashes, 35.7% had occurred due to a rear-end collision and 23.9% had occurred due to a lane-change maneuver (8.9% and 6.0% for all crashes respectively). In 31.6% of cases, the truck was not involved in the initial conflict e.g. was the third vehicle in a rear end collision.

Commercial vehicles are the second most frequent crash opponents of 16t+ trucks (22.9%). Cases where two 16t+ trucks were involved in a crash are counted from each participant's perspective. Accidents in longitudinal traffic are again the most frequent types (76.8%, 17.6% of all cases). In 28.1% of longitudinal cases, the 16t+ truck is the striking vehicle in a rear-end collision, and in 35.4% the struck vehicle. In contrast to accidents with cars, lane change accidents rarely occurred between a heavy truck and another commercial vehicle.

Among the vulnerable road users, cyclists have the highest share as crash opponents for 16t+ trucks. The most common specific accident type with 44.8% is when a truck turns right and has a conflict with a cyclist travelling alongside in the same direction mostly on a bicycle path on the right side of the road.

Table 2. Categories of accident types for 16t+ trucks with different types of road users as crash opponents, based on GIDAS

| Categories based on initial conflict | Cars | Commercial Vehicles | Bicycles | Pedestrian | Powered Two-Wheeler | *TOTAL* |
|---|---|---|---|---|---|---|
| 1 Driving accident | 39 | 12 | 3 | 0 | 3 | *57* |
| 2 Turning off accident | 44 | 11 | 50 | 10 | 10 | *125* |
| 3 Crossing/entering accident | 81 | 15 | 30 | 0 | 7 | *133* |
| 4 Pedestrian crossing road | 1 | 0 | 0 | 34 | 0 | *35* |
| 5 Accident with parked vehicle | 20 | 8 | 4 | 2 | 3 | *37* |
| 6 Accident in longitudinal traffic | 272 | 192 | 8 | 2 | 14 | *488* |
| 7 Other accident types | 30 | 12 | 1 | 8 | 0 | *51* |
| *TOTAL* | *487* | *250* | *96* | *56* | *37* | *926* |

Conflicts between 16t+ trucks and pedestrians were found in 5.1% of all cases. The most frequent category among those crashes is when a pedestrian enters the road to cut across perpendicular to the direction of travel of the truck in 60.7% of cases. The second most frequent category of crashes with pedestrians is when the truck turned off the main road and had a conflict with a pedestrian walking on the side walk (17.9%).

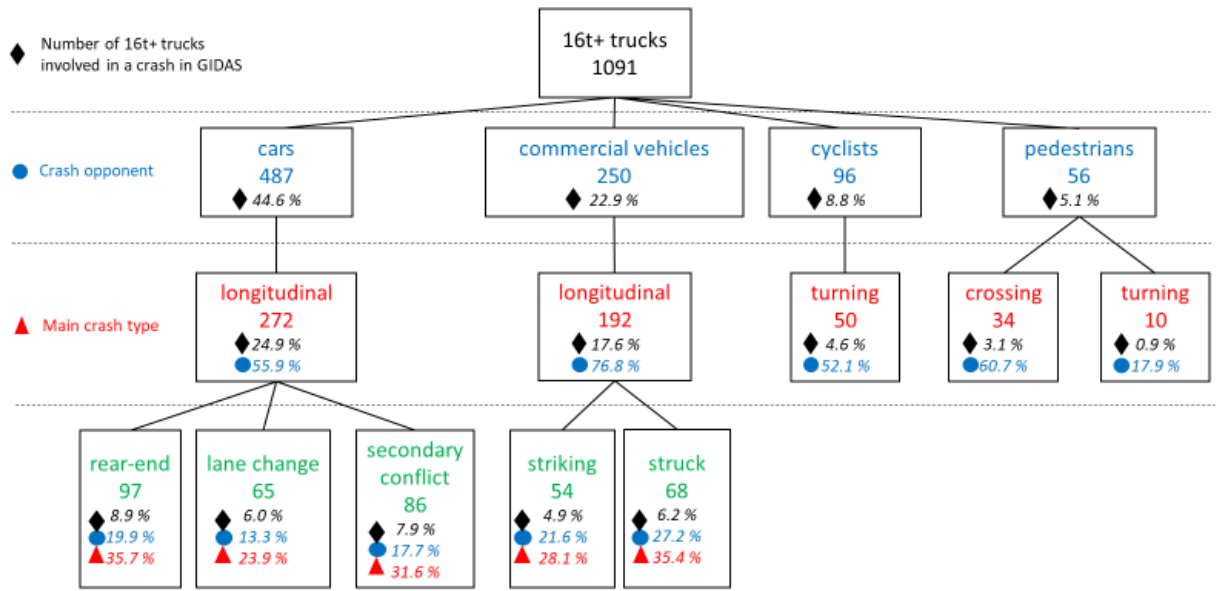

Fig. 3 Most common crash types from GIDAS (case count and percentages per category)

Crashes of 16t+ trucks with powered two wheelers were not as common as with other road users (3.4% of cases). These cases were mostly from the accident type categories of "turning off accidents" or "accidents in longitudinal traffic" and are not further investigated due to the low number of cases.

Based on the most common accident types and results from CARE and national crash databases, the following crash scenarios for 16t+ trucks were selected for further analysis in the GIDAS dataset:

- Scenario 1: rear-end crashes with cars and commercial vehicles as collision partners. Due to the focus on preventability of the crash, cases where the 16t+ truck is the striking vehicle are considered.
- Scenario 2: conflicts between a truck that is turning to the right and a cyclist that is travelling alongside with the intention to go straight.
- Scenario 3: conflicts between pedestrians crossing the road and trucks.

*3.3.1. Scenario 1: Rear-end crashes with 16t+ truck as striking vehicle*

In the first scenario, consisting of 219 cases in the GIDAS database (20% of all cases), the average travelling speed of the trucks at conflict initiation was 50km/h, with 25% of the trucks travelling at speeds above 80km/h (Fig. 4). The initial speeds of the struck vehicles were substantially lower at the initiation of the conflict at an average of 20km/h. The analysis of the collision speeds showed that truck drivers often had the time to initiate a braking maneuver as collision speeds were lower than the initial travelling speeds (see Fig. 4). On the other hand, when struck, the collision opponent of this scenario had been standing in most cases.

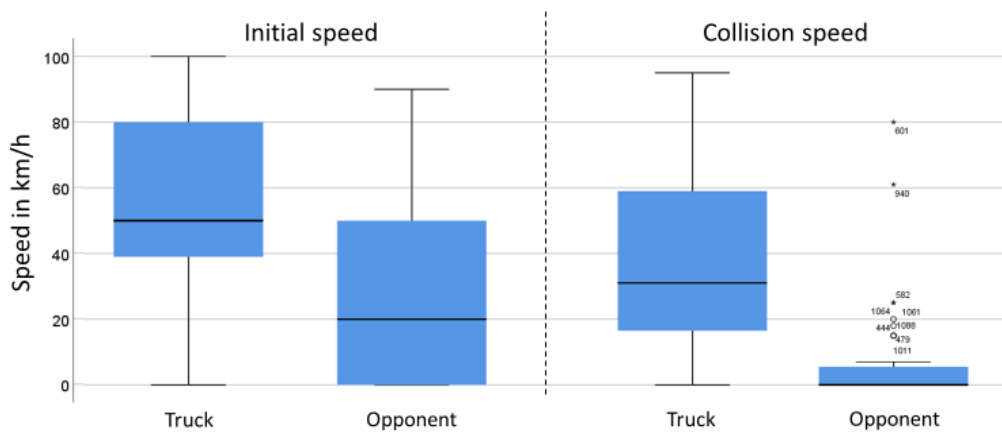

Fig. 4 Scenario 1, initial speeds of truck and collision partner in rear end crashes, based on GIDAS

*3.3.2. Scenario: Right-turn conflicts with cyclists*

The second scenario includes right turning trucks that had a conflict with a cyclist travelling alongside along the initial direction of the truck in 43 cases (4% of all cases) and only one case where the cyclist was going in the opposite direction. The conflicts occurred at lower collision speeds of the truck compared with scenario 1, with an average of 13km/h (see Fig. 5). No reliable speed data is available for the cyclist.

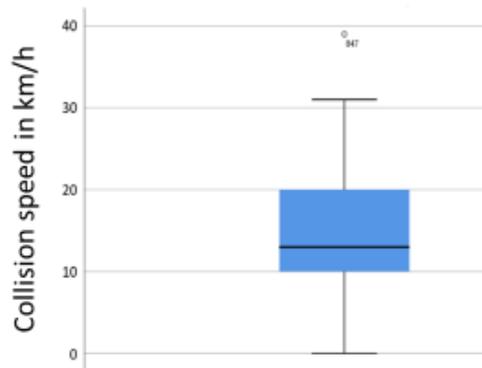

Fig. 5 Scenario 2, collision speed of truck in right turn crashes, based on GIDAS

The collision angle between cyclist and truck describes the angle between the motion vectors of the truck and cyclist at the point of the collision. It was found to be between 0° and 60° in most cases with a peak at 30° (see Fig. 6). In 75% of the cases, the cyclist had collided with the side of the truck in the range of 2m from the front of the truck. Only 4 contact points had been further away from the front than 5m (i.e. behind the cabin and further towards the trailer/rear axle of the truck).

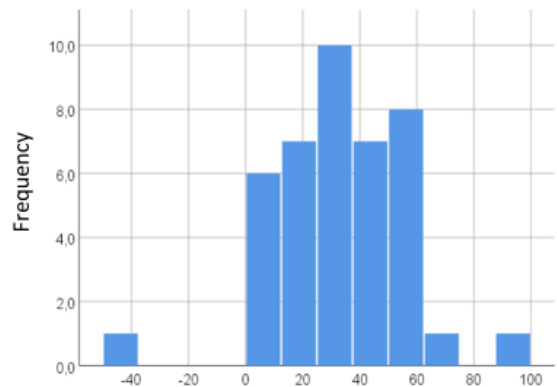

Fig. 6 Scenario 2, collision angles between truck and cyclist in right turn crashes, based on GIDAS

*3.3.3. Scenario 3: Pedestrian crossing conflicts*

In the third scenario, consisting of 34 cases (3% of all cases), the pedestrian was overrun by a part of the truck in 16 cases (which leads to more severe or fatal injuries). In 10 out of the 16 cases, the truck had initially been standing (e.g. at a traffic light) and the pedestrian had crossed the road directly in front of the truck when the truck started to accelerate. Thus, the collision speeds here were under 10km/h in most cases (see Fig. 7). In the remaining cases where the pedestrians were not overrun, the truck had a considerably higher average collision speed at 23km/h (in 25% of cases even above 40km/h).
The speed at the initiation of the conflict when the pedestrian was not overrun were similar to the collision speed (see Fig. 7), suggesting that in many cases there was little time available for the truck driver to react and brake.

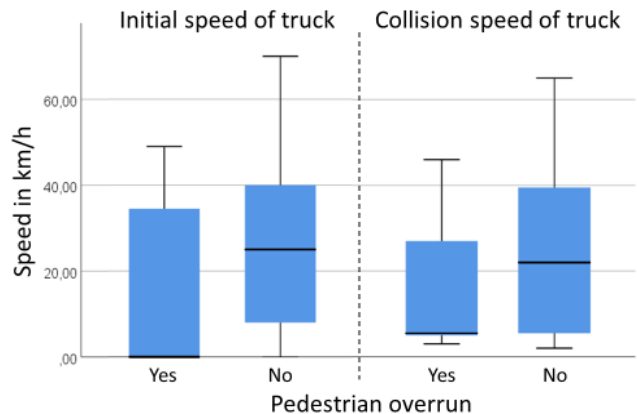

Fig. 7 Scenario 3, initial and collision speeds of truck in pedestrian crossing crashes, based on GIDAS

## 4. Discussion

Differences seen between the data from CARE, national databases and GIDAS can result from the existence of local differences (e.g. higher exposure to rainy weather conditions in Sweden than in Spain) or different filter criteria (i.e. weight restriction to above 3.5t for CARE and 16t for national databases and GIDAS). Further uncertainties are introduced by the coding scheme. Different countries are using different coding schemes, e.g. for crash types. Certain variables might not be reported at all in some databases (e.g. day/nighttime in Italian data).
For this analysis of national crash data, crash types have been re-coded to a common scheme, following the Swedish crash type categorization. There are however crash types that are difficult to re-code, and even at the point of data collection, police officers might have difficulties to categorize the crash. Indicators for the difficulty can be seen in single vehicle crashes, where by definition only truck occupants should be injured, but also car occupant and VRU injuries are sometimes reported. Furthermore, depending on the variable, from 0.5% to 20% of injuries can be reported as unknown, either as no information is available or uninjured persons will possibly be classified as unknown. Removing "unknown" values from the analysis could potentially reduce the number of reported cases for some categories more than for others and therefore introduce a bias into the relations between categories.
The results obtained in the crash data analysis at hand supports the findings of similar previous studies in the US, such as Zhu and Srinivasan (2011) who identified collisions in longitudinal traffic and collisions at intersections as the most common crash types. Kockum et al. (2017) had identified cars and other HGVs as the most common collision partners, and these findings are supported by the outcomes of the analysis at hand. The results of this study support the existing knowledge and extend it to provide a detailed picture of HGV crashes in Europe.

## 5. Conclusion

This study provides a deeper understanding of crashes involving 16t+ trucks in Europe by a comprehensive data analysis conducted simultaneously on three levels. Most crashes occur during dry and clear weather conditions (76%-88%, depending on region), during daylight conditions (73%-78%) and on dry roads (51%-83%), outside city limits (57%-87%) on non-highway roads (54%-81%). All three analysis levels show the same trends regarding these variables, but small differences exist. The reasons for these differences could range from local effects (e.g. weather, driving behavior, vehicle types) to filter criteria in each database (e.g. weight or size restrictions).
As a result of the three-stage analysis, three scenarios were identified that should be addressed by future research and safety systems: (1) rear-end crashes with other vehicles in which the truck is the striking partner, (2) conflicts during right turn maneuvers of the truck and a cyclist that is travelling alongside with the intention to go straight and (3) pedestrians crossing the road perpendicular to the direction of travel of the truck. These three scenarios were studied in more detail in the GIDAS database, leading to the following conclusions:
(1) In rear-end crashes, the average speed of the truck at conflict onset is about 50km/h and at the time of the collision, it is reduced to about 30km/h, whereas the struck vehicle is typically standing at impact.
(2) During the right turn maneuvers, the average collision speed of the truck is about 13km/h and the impact happens at an angle of 33 degree on average, with the impact point within the first 2 m along the length of the truck (i.e. around the area of the passenger-side door).
(3) In the pedestrian crossing scenario, speeds are generally low (< 5km/h) for cases where the pedestrian is overrun by the truck, often resulting from situations when pedestrians cross in front of a waiting truck that started to accelerate. Collision speeds are higher (>20km/h) when the pedestrians are not overrun. Although the speeds in the VRU scenarios are generally low, the outcomes are severe, especially when the VRUs are overrun by the truck.

## Acknowledgements

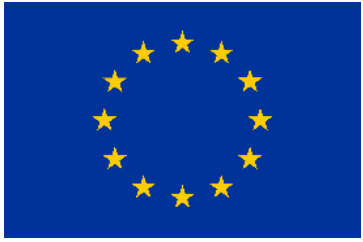

This project has received funding from the European Union's Horizon 2020 research and innovation program under grant agreement No 769658. The contents of this publication reflect only the author's view and the European Agency is not responsible for any use that may be made of the information it contains. The authors would like to thank Núria Parera (IDIADA, Spain) and Giuseppe Cordua (IVECO, Italy), for their valuable input and contributions for the analysis of the respective national databases. The data collection and processing for the GIDAS project is funded by the Federal Highway Research Institute (BASt) and the German Research Association for Automotive Technology (FAT), a department of the VDA (German Association of the Automotive Industry).